\documentclass[a4paper,12pt]{article}

\usepackage [T1]{fontenc}
\usepackage {calligra}
\usepackage [T1]{pbsi}
\usepackage{CJKutf8} 
\usepackage[colorlinks=true,linkcolor=blue]{hyperref}
\usepackage{amsmath,amssymb,amsfonts} 
\usepackage{graphicx}                 
\usepackage{color}                    
\usepackage{hyperref}                 
\usepackage{cite}
\definecolor{red}{rgb}{1,0,0}           
\definecolor{green}{rgb}{0,1,0}
\definecolor{blue}{rgb}{0,0,1}
\definecolor{darkblue}{rgb}{0,0,0.5}
\definecolor{lightblue}{rgb}{.5,.5,1}
\definecolor{lightgray}{gray}{.87}          
\definecolor{Dark}{gray}{.20}
\definecolor{pink}{rgb}{.95,0.82,0.92}  
\definecolor{yellow}{rgb}{1,1,0}
\definecolor{lightyellow}{rgb}{1,1,.5}
\definecolor{purple}{rgb}{0.7,0,0.85}
\definecolor{darkgreen}{rgb}{0,0.45,0}
\definecolor{orange}{rgb}{0.8,0.2,0.2}
\def \be {\begin{equation}}
\def \ee {\end{equation}}
\def \bea {\begin{eqnarray}}
\def \eea {\end{eqnarray}}
\def \nn {\nonumber}

\def \rr {\raise.35ex\hbox{\small $\prime$}\kern-.17em{\mbox{\large $\imath$}}}
\def \del {\partial}
\def \dels {\partial\kern-.5em / \kern.5em}
\def \As {{A\kern-.5em / \kern.5em}}
\def \Ds {D\kern-.7em / \kern.5em}

\def \a {\alpha}

\def \b {\beta}

\def \g {\gamma}
\def \G {\Gamma}
\def \d {\delta}
\def \eps {\epsilon}
\def \m {\mu}
\def \n {\nu}

\def \lam {\lambda}
\def \Lam {\Lambda}
\def \s {\sigma}
\def \r {\rho}
\def \om {\omega}

\def \th {\theta}

\def \teps {\tilde{\epsilon}}

\newcommand{\solution}[1]{}

\newcommand{\hide}[1]{}

\begin{document}

\pagestyle{plain}

\begin{CJK}{UTF8}{bsmi} 

\begin{titlepage}


\begin{center}

\noindent
\textbf{\LARGE
Generalized Yang-Mills Theory and Gravity
\vskip.6cm
}

\vskip .5in
{\large 
Pei-Ming Ho$ $\footnote{e-mail address: pmho@phys.ntu.edu.tw},
}
\\

{\vskip 10mm \sl
Department of Physics and Center for Theoretical Sciences \\
Center for Advanced Study in Theoretical Sciences \\
National Taiwan University, Taipei 106, Taiwan,
R.O.C.
}\\
\vskip 3mm
\vspace{60pt}

\begin{abstract}

We propose a generalization of Yang-Mills theory for which
the symmetry algebra does not have to be factorized as
mutually commuting algebras of a finite-dimensional Lie algebra 
and the algebra of functions on base space.
The algebra of diffeomorphism can be constructed as an example,
and a class of gravity theories can be interpreted as generalized Yang-Mills theories.
These theories in general include a graviton, a dilaton 
and a rank-2 antisymmetric field, 
although Einstein gravity is also included as a special case.
We present calculations suggesting that
the connection in scattering amplitudes between 
Yang-Mills theory and gravity via BCJ duality
can be made more manifest in this formulation.

\end{abstract}

\end{center}

\end{titlepage}

\setcounter{page}{0}
\setcounter{footnote}{0}
\setcounter{section}{0}


\section{Introduction}

The textbook definition of Yang-Mills (YM) theory 
is usually based on the choice of a finite dimensional Lie group,
and gauge transformations are specified by Lie-group valued functions.
However, 
it is well known that 
when the base space is noncommutative,
the algebra of gauge transformations 
is a mixture of the finite-dimensional Lie algebra
and the algebra of functions on the noncommutative space.
As a result,
$SU(N)$ gauge symmetry cannot be 
straightforwardly defined on noncommutative space.

In this paper, 
we consider a minor generalization
of the notion of gauge symmetry.
We will not only allow the generators of gauge transformations 
to behave like pseudo-differential operators
(as functions on noncommutative space do),
but we will also allow them
to be not factorized into the part of a finite-dimensional Lie algebra
and that of functions on the base space.
That is,
the gauge symmetry algebra does not have to be defined 
as the composition of a finite-dimensional Lie algebra 
and an associative algebra of functions on the base space.
With this generalization,
it may no longer be possible to view a gauge symmetry 
as what you get from ``gauging'' a global symmetry
through introducing space-time dependence.

A possibility of this generalization was already suggested \cite{Ho:2001as}
for even-dimensional spherical brane configurations in the matrix theory.
For example, 
for the fuzzy-$S^4$ configuration of $n$ D4-branes,
the algebra of functions on the 4-dimensional base space is non-associative,
but there is an associative algebra for gauge transformations.
For large $n$,
the gauge symmetry algebra is approximately 
that of a $U(n)$-bundle 
(or equivalently a fuzzy-$S^2$ bundle) 
over $S^4$ \cite{Ho:2001as}.

Another example is the low energy effective theory of a D3-brane 
in large R-R 2-form field background \cite{Ho:2013opa}.
This theory is S-dual to the noncommutative gauge theory 
for a D3-brane in large NS-NS $B$-field background.
The gauge symmetry to all orders in the dual theory 
is not given by the noncommutative gauge symmetry,
but is characterized by a bracket $\{\cdot, \cdot\}_{**}$
which defines a non-associative algebra on the base space \cite{Ho:2013opa}.
(The gauge symmetry algebra is of course associative.)

In this paper,
we will show that the gauge symmetry of space-time diffeomorphism 
is also an example of the generalized gauge symmetry.
Accordingly,
a class of gravity theories can be interpreted as YM theories.
Generically, 
these theories include a graviton, a dilaton and an anti-symmetric tensor.
We will point out that the connection between Yang-Mills theory and gravity
(through the color-kinematics duality)
is manifest at tree level in 3-point amplitudes.

Attempts to interpret gravity as a gauge theory have a long history
since the works of Utiyama \cite{Utiyama}, Kibble \cite{Kibble} and Sciama \cite{Sciama}.
It is well known that General Relativity (GR) can be rewritten 
as the Chern-Simons theory in 3 dimensions \cite{Witten:1988hc},
and a YM-like theory in 4 dimensions \cite{MacDowell:1977jt,Stelle-West},
as well as higher dimensions \cite{Vasiliev:2001wa}.
The vielbein and the connection are defined
as components of a gauge potential,
and the gauge symmetry is $SO(d,2)$,
instead of the space-time diffeomorphism.
These formulations are based on gauge symmetries 
in the traditional sense.
Our formulation of gravity as a YM theory is 
different from these formulations.

While GR can be formulated as a YM theory,
YM theories can also be realized as the low energy effective theories 
of gravity theories in higher dimensions via suitable compactification.
Similar to this scenario of Kaluza-Klein reduction, 
internal symmetries and external symmetries are 
treated on equal footing in the generalized YM theories,
as we will not distinguish the base space dependence
from the internal space dependence in the gauge symmetry algebra.

Our formulation of gravity is also reminiscent of teleparallel gravity \cite{review-tele},
which can be interpreted as a gauge theory of
the (Abelian) translation group,
with the vielbeins playing the role of the gauge potential.
In another formulation of gravity \cite{Cortese:2010ze}
in which the vielbeins are identified with the gauge potential,
a deformation of the gauge symmetry is considered 
to achieve the nonlinearity in gravity.
In our formulation,
on the other hand,
the gauge potential is not the vielbein, 
but the inverse of the vielbein.

The plan of this paper is as follows.
In Sec.\ref{GaugeSymm},
we will see how the algebra of diffeomorphism 
appears as an example of the generalized gauge symmetry.
In Sec.\ref{YM},
the gauge potential for the gauge symmetry of diffeomorphism 
is essentially the inverse of the vielbein, 
and the field strength is the torsion of the Weitzenb\"{o}ck connection.
We show that the corresponding YM theories
with quadratic Lagrangians 
define a class of gravity theories in Sec.\ref{YM=GR}.
It will be pointed out in Sec.\ref{ScattAmp}
that this new formulation of gravity 
may have significant advantages in its use to 
compute scattering amplitudes,
with relations reminiscent of the double-copy procedure \cite{BCJ}
to derive scattering amplitudes in gravity 
from YM theories.
In Sec.\ref{HigherGauge},
we comment on extensions of the generalized notion 
of gauge symmetry to higher form gauge theories.

\section{Gauge Symmetry Algebra}
\label{GaugeSymm}

In a naive textbook introduction to non-Abelian gauge symmetry,
the gauge transformation parameter
$\Lam(x) = \sum_a \Lam^a(x)T_a$
is a sum of products of space-time functions and Lie algebra generators.
The Lie algebra of local gauge transformations is 
spanned by a set of basis elements, say,
\be
T_a(p) \equiv e^{ip\cdot x} T_a
\label{factor}
\ee
in the Fourier basis.
In this basis,
a gauge transformation parameter can be expressed as
\be
\Lam(x) = \sum_{a, p} \tilde{\Lambda}^a(p) T_a(p),
\ee
where the sum over $p$ is understood to be the integral $\int d^D p$
for $D$-dimensional space-time.
Similarly, 
the gauge potential 
can be written as
\be
A_{\mu}(x) = \sum_{a, p} \tilde{A}^a_{\mu}(p) T_a(p).
\ee

Normally,
for a given finite-dimensional Lie algebra with structure constants $f_{ab}{}^c$,
the algebra of gauge transformations has the commutator
\be
[T_a(p), T_b(p')] = \sum_c f_{ab}{}^c T_c(p+p'),
\ee
where the structure constants $f_{ab}{}^c$ only involve color indices $a, b, c$.
For these cases, 
the inclusion of functional dependence on the space-time in the generators $T_a(p)$ is trivial,
and thus often omitted in discussions.

However, 
for noncommutative gauge symmetries,
the structure constants depend not only on the color indices $a, b, c$,
but also on the kinematic parameters $p, p'$.
For a noncommutative space defined by
\be
[x^{\mu}, x^{\nu}] = i\theta^{\mu\nu},
\ee
the Lie algebra $U(N)$ gauge symmetry is
\be
[T_a(p), T_b(p')] 
= \sum_{p''} {\bf f}_{ab}{}^{c}(p, p', p'') T_c(p''),
\label{NCUN}
\ee
where the structure constants are
\footnote{
Here $p\th p'$ stands for $p_{\m}\th^{\m\n}p'_{\n}$.
}
\be
{\bf f}_{ab}{}^{c}(p, p', p'') =
\left[
f_{ab}{}^c\left(e^{ip\th p'} - e^{ip'\th p}\right)
+ d_{ab}{}^c\left(e^{ip\th p'} + e^{ip'\th p}\right)
\right]\d^{(D)}(p+p'-p''),
\ee
and they involve kinematic parameters $p, p'$ and $p''$.
Here $f_{ab}{}^c$ is the structure constant of $U(N)$
and $d_{ab}{}^c$ is defined by
\be
\{T_a, T_b\} = d_{ab}{}^c T_c
\ee
for $T_a$'s in the fundamental representation.
In this gauge symmetry algebra,
the $U(N)$ Lie algebra and the algebra of functions on the base space are mixed.
(This is the obstacle to define noncommutative $SU(N)$ gauge symmetry.)
The gauge algebra is non-Abelian even for the Abelian group $U(1)$.

To describe the noncommutative $U(N)$ gauge algebra properly,
it is a necessity to use the generators (\ref{factor}) including 
functional dependence on the base space.
Nevertheless, 
the noncommutative $U(N)$ gauge symmetry still assumes 
that the generators can be factorized (\ref{factor}),
and that $e^{ip\cdot x}$ always commutes with $T_a$.
These are unnecessary assumptions for 
most algebraic calculations in the gauge theory.
After all, in field theories,
only the coefficients $\tilde{A}^a_{\mu}(p)$ are operators (observables),
while the space-time and Lie algebra dependence
are to be integrated out (summed over) in the action.

It is thus natural to slightly extend 
the formulation of gauge symmetry
(and YM theory)
to allow the Lie algebra to be directly defined 
in terms of the generators $T_a(p)$,
without even assuming its factorization into 
a Lie algebra factor $T_a$ and a function $e^{p\cdot x}$.
The integration over space-time and trace of the internal space
in the action can be replaced by the Killing form
of the Lie algebra of $T_a(p)$.
The distinction between internal space and external space
is reduced in this description.

In short, 
we propose to study 
gauge symmetries 
without assuming its factorization into two associative algebras
(an algebra for the functions on the base space and 
a finite dimensional Lie algebra).
Even when it is possible to factorize the generators formally as (\ref{factor}), 
we will not assume that the space-time functions 
to commute with the algebraic elements $T_a$.
(In general,
we do not have to use the Fourier basis,
and the argument $p$ of $T_a(p)$ can represent labels 
of any complete basis of functions on the base space.)
One of the goals of this paper is to show that 
this generalization is beneficial, 
for bringing in new insights into gravity theories.

Algebraically,
this generalization is very natural.
A corresponding geometric notion is however absent at this moment.
(It is not clear what {\em twisted} bundles would mean.)
The notion of bundles on noncommutative space
is replaced by projective modules \cite{Connes},
which should be further generalized for our purpose.
We postpone the problem of finding a suitable geometric notion
for the generalized gauge symmetry to the future.

\subsection{A Generalized Gauge Symmetry Algebra}

For the generalized gauge symmetry,
we consider a Lie algebra defined as
\be
[T_a(\a), T_b(\b)] = \sum_{\g} \; f_{ab}{}^c(\a, \b, \g) T_c(\g),
\ee
which may not be decomposed as a product of 
the algebra of functions on the space-time
and a finite-dimensional Lie algebra.
Here $\a, \b, \g$ are labels for a complete basis of functions 
on the base space.
For a theory with translational symmetry, 
it would be natural to use the Fourier basis, 
and we have
\be
[T_a(p), T_b(p')] = f_{ab}{}^c(p, p') T_c(p+p'),
\ee
where the structure constant is actually $f_{ab}{}^c(p, p') \d^{(D)}(p+p'-p'')$
with the Dirac delta function cancelled by the integration over $p''$.

The Jacobi identity of this Lie algebra is
\be
f_{ab}{}^e(p, p')f_{ec}{}^d(p+p', p'') + 
f_{bc}{}^e(p', p'')f_{ea}{}^d(p'+p'', p) + 
f_{ca}{}^e(p'', p)f_{eb}{}^d(p''+p, p') = 0.
\ee
Every solution to this equation for $f$ defines a gauge symmetry.
It will be interesting to find solutions with non-trivial dependence on momenta,
as the case of the noncommutative gauge symmetry.

As an example,
let us now construct a Lie algebra for generators of the form
\be
\{ T(\teps, p) \equiv \teps^{(\mu)}T_{(\mu)}(p) \},
\ee
where the basis elements $T_{(\mu)}(p)$ has 
a space-time index $\mu$ as its internal space index.
To construct a concrete example,
we will assume that the structure constants are linear in momenta $p$, $p'$,
and that it is compatible with Poincare symmetry.

We put the index $\mu$ in a parenthesis 
to remind ourselves that it plays the role of 
the index of an internal space.
The reason why we choose this ansatz for a Lie algebra 
is that the Lie algebra of space-time diffeomorphisms is of this form.
We wish to explore the possibility of 
rewriting a gravity theory as a generalized YM theory.

The translation symmetry implies that 
the commutators are of the form
\be
[T(\teps_1, p_1), T(\teps_2, p_2)] 
= \sum_{\teps_3} f(\teps_1, p_1, \teps_2, p_2, \teps_3)
T(\teps_3, p_1 + p_2),
\label{ansatz}
\ee
and the Lorentz symmetry implies that
the structure constants are Lorentz-invariant functions
of the vectors $\teps_1, \teps_2, \teps_3, p_1, p_2$.
(The summation over $\tilde{\eps}_3$ 
is a summation over an orthonormal basis of vectors.)
The most general ansatz consistent with all assumptions is thus
\be
f(\teps_1, p_1, \teps_2, p_2, \teps_3) =
(\teps_1\cdot\teps_3) 
\left[\teps_2\cdot (\a p_1 + \g p_2) \right]
- (\teps_2\cdot\teps_3) 
\left[\teps_1\cdot (\a p_2 + \g p_1) \right]
+ \b (\teps_1\cdot\teps_2) \left[\teps_3\cdot (p_1 - p_2)\right],
\ee
where $\a, \b, \g$ are constant parameters.

It follows from the ansatz that
\begin{align}
&[[T(1), T(2)], T(3)] + [[T(2), T(3)], T(1)] + [[T(3), T(1)], T(2)]
= \sum_{\teps_4} \Big\{
\nn \\
&(\teps_1\cdot\teps_4)\left[
- \a\b (\teps_2\cdot\teps_3)(p_1\cdot(p_2-p_3))
+ \g [(\teps_2\cdot p_2)(\teps_3\cdot (\a p_1 + \g p_2)
- (\teps_3\cdot p_3)(\teps_2\cdot (\a p_1 + \g p_3)]
\right]
\nn \\
&+ \b (\teps_1\cdot\teps_2)\left[
(\teps_3\cdot(p_1 - p_2))
(\teps_4\cdot[(\a-\b)(p_1 + p_2) - \b(p_1 - p_2)])
- \g (\teps_3\cdot p_3)(\teps_4\cdot(p_1 - p_2))
\right]
\nn \\
&+ \mbox{cyclic permutations of $(1, 2, 3)$}
\Big\},
\end{align}
where we have used $T(1)$ to represent $T(\teps_1, p_1)$,
$T(2)$ to represent $T(\teps_2, p_2)$, etc.
In order to satisfy the Jacobi identity,
we have to set $\b = \g = 0$.
The most general solution is thus equivalent to
\be
[T(1), T(2)] = 
\sum_{\teps_3}
i \left[
(\teps_2\cdot\teps_3) (\teps_1\cdot p_2)
- (\teps_1\cdot\teps_3) (\teps_2\cdot p_1)
\right] T(3),
\label{commutator}
\ee
by scaling $\a$ to $-i$.
More explicitly,
it is
\be
[T_{(\mu)}(p_1), T_{(\nu)}(p_2)] = \sum_{\lam}
i\left[
\eta_{\nu\lam}p_{2\mu} - \eta_{\mu\lam}p_{1\nu}
\right] T_{(\lam)}(p_1+p_2).
\label{commutator-2}
\ee

Incidentally, 
it is consistent to allow $\a$ to depend on the momenta.
For example, 
Jacobi identity is satisfied for
\be
\a(p_1, p_2) = c \, e^{\lam p_1\cdot p_2}
\ee
for arbitrary constant parameters $c$ and $\lam$.
It is equivalent to the scaling of the generators by
$T(\eps, p) \rightarrow T'(\eps, p) \equiv c\, e^{\lam p^2/2} T(\eps, p)$.

\subsection{Representations}

A representation of the algebra constructed above is given by 
\be
T(\teps, p) \tilde{\phi}(p') = i (\teps\cdot p') \tilde{\phi}(p'+p),
\ee
on a linear space with the basis $\{\tilde{\phi}(p)\}$.
This expression allows us to interpret $\tilde{\phi}(p)$
as the Fourier modes of a scalar field $\phi(x)$
and $T(\teps, p)$ as the generator of 
a coordinate transformation with 
\be
\d x^{\mu} = \teps^{(\m)} e^{- i p\cdot x}.
\ee
That is,
\be
T(\teps, p) = e^{ip_{\nu} x^{\nu}} \teps^{(\m)}\del_{\m}.
\ee
Indices are contracted according to Einstein's summation convention 
regardless of whether they are in parentheses or not.
In this representation, 
it is clear that this algebra is that of space-time diffeomorphism.
The gauge symmetry of GR arises as the most general gauge symmetry 
with Lie algebra of the form (\ref{ansatz}),
assuming that the structure constants are linear in momentum
and that they respect Poincare symmetry.

A generic element of the Lie algebra is a superposition
\be
\int d^D p \; \teps^{(\m)}(p) T_{(\m)}(p)
\ee
in $D$ dimensions,
which can be written as
\be
T_{\eps} \equiv \eps^{(\m)}(x) \del_{\mu},
\label{Teps}
\ee
where $\eps^{(\m)}(x)$ is the inverse Fourier transform of $\teps^{(m)}(p)$.


In view of this representation (\ref{Teps}),
it is tempting to interpret the algebra constructed above as
merely the result of taking $T_a$'s to be derivatives $\del_a$'s in (\ref{factor})
for a traditional gauge symmetry.
But if we were really dealing with a traditional gauge symmetry,
we would have obtained an Abelian gauge symmetry because
$[T_a, T_b] = [\del_a, \del_b] = 0$.
The need to generalize the notion of gauge symmetry here is
due to the fact that $T_a = \del_a$ does not commute with space-time functions.
Incidentally, 
the traditional interpretation of the torsion in teleparallel gravity
is indeed the field strength of an Abelian gauge theory \cite{Cho:1975dh}.
(See eq.(\ref{torsion-field-strength}) below.) 

Matter fields in the gauge theory are classified as
representations of the gauge symmetry.
Since the gauge symmetry under consideration is the diffeomorphism, 
we know all about other representations of different spins.

\section{Gauge Field of Diffeomorphism}
\label{YM}

\subsection{Gauge Potential vs Vielbein}
\label{vielbein}

We can define a gauge potential for the gauge symmetry algebra (\ref{commutator}).
In the representation (\ref{Teps}),
the gauge potential 
\be
A_{\mu}(x) = \sum_{\nu, p} \tilde{A}_{\mu}{}^{(\nu)}(p)T_{(\nu)}(p)
= A_{\mu}{}^{(\nu)}(x) \del_{\nu}
\ee
should transform like
\be
\d A_{\mu}(x) = [D_{\mu}, \Lam(x)],
\label{A-transf}
\ee
where 
\be
D_{\mu} \equiv \partial_{\mu} + A_{\mu}(x) = (\d_{\mu}^{\nu} + A_{\mu}{}^{(\nu)})\del_{\nu}
\ee
is the covariant derivative 
and the gauge transformation parameter is
\be
\Lam(x) = \Lam^{(\mu)}(x) \del_{\mu}.
\ee

More explicitly, 
the gauge transformation (\ref{A-transf})
can be expressed as
\be
\d A_{\mu}{}^{(\nu)}(x)
= \del_{\mu} \Lam^{(\nu)}(x)
- \Lam^{(\lam)}(x) \del_{\lam} A_{\mu}{}^{(\nu)}(x)
+ A_{\mu}{}^{(\lam)}(x) \del_{\lam} \Lam^{(\nu)}(x).
\label{Amn-transf}
\ee

Let us recall that the vielbein $e_{\mu}{}^a(x)$ in gravity 
is defined to transform under general coordinate transformations as
\be
\d e_{\mu}{}^a(x) = 
\d x^{\nu} \del_{\nu} e_{\mu}{}^a 
+ e_{\nu}{}^a \del_{\mu}\d x^{\nu}.
\ee
The index $a$ on $e_{\mu}{}^a$ labels a local orthonormal Lorentz frame.
Under a rotation of the local Lorentz frame,
\be
e_{\m}{}^a(x) \rightarrow e'_{\m}{}^a(x) = \om^a{}_b(x) e_{\m}{}^b(x),
\label{rotation-0}
\ee
where
\be
\om_{ab}(x) = - \om_{ba}(x)
\ee
is the parameter for infinitesimal $SO(D-1, 1)$-rotations.

The inverse $e_a{}^{\mu}(x)$ of the vielbein is defined by
\be
e_a{}^{\mu}(x) e_{\mu}{}^b(x) = \d_a^b, 
\qquad
e_{\mu}{}^a(x) e_a{}^{\nu}(x) = \d_{\mu}^{\nu}.
\ee
The transformation of $e_a{}^{\mu}(x)$ is
\be
\d e_a{}^{\mu}(x) =
\d x^{\nu} \del_{\nu} e_a{}^{\mu}(x)
- e_a{}^{\nu} \del_{\nu} \d x^{\mu}.
\label{e-inv-transf}
\ee

Let us now consider a flat background in which 
\be
e_a{}^{\mu}(x) = \d_a^{\mu} + C_a{}^{\mu}(x)
\ee
for a fluctuation denoted by $C_a{}^{\m}$.
Here we have chosen a particular frame in which 
the flat background is given by $e_a{}^{\mu} = \d_a^{\mu}$.
The local Lorentz transformation symmetry is not manifest
(nonlinearly realized)
in terms of the variable $C_a{}^{\mu}$.

It follows from (\ref{e-inv-transf}) that
\be
\d C_a{}^{\mu}(x) = 
- \del_{a} \d x^{\mu} +
\d x^{\nu} \del_{\nu} C_a{}^{\mu}(x)
- C_a{}^{\nu} \del_{\nu} \d x^{\mu}.
\ee
Comparing this expression with (\ref{Amn-transf}),
we see that it is tempting to identify 
$A$ with $C$, and $\Lam$ with $- \d x$.

The transformations of $A$ and $C$ are matched 
by identifying upper (lower) indices with upper (lower) indices.
That is, 
the Lorentz index $\mu$ 
on $A_{\mu}{}^{(\nu)}$ is to be identified with 
the local Lorentz frame index $a$
on $C_a{}^{\mu}$,
while the internal space index $\nu$ 
on on $A_{\mu}{}^{(\nu)}$ is to be identified with 
the space-time coordinate index $\mu$ on $C_a{}^{\mu}$.
This may seem peculiar at first sight but it is actually expected.
The gauge algebra (\ref{commutator}) is defined 
with the assumption of Poincare symmetry on the base space,
so the Lorentz index $\mu$ on the potential $A_{\mu}{}^{(\nu)}$
cannot be identified with the coordinate index $\mu$
on a curved manifold.
On the other hand,
the internal space index $\nu$ on the potential $A_{\mu}{}^{(\nu)}$ 
is contracted with the index of a derivative $\del_{\nu}$ of space-time coordinates,
hence it is really an index of coordinates.

Note that the gauge potential $A$ is still defined
as part of the covariant derivative
$D = dx^{\mu} D_{\mu}$,
and in this sense it is still a 1-form.
The one-form index $\mu$ of $A_{\mu}{}^{(\nu)}$
is matched with the frame index,
not the 1-form index of the vielbein,
only because the geometric interpretation of gravity is changed.
The gauge symmetry of gravity is now interpreted
as a non-Abelian symmetry on Minkowski space
whose transformations involve kinematic vectors.
On the other hand,
the potential $A_{\mu}{}^{(\nu)}$ is not a pure 1-form
as it has a vector-field index $(\nu)$.

In the following,
we will adopt the conventional notation for vielbeins.
Latin letters $a, b, c, \cdots$ are used for indices of local Lorentz frames,
and Greek letters $\mu, \nu, \lam, \cdots$ for indices of space-time coordinates.
For instance, 
we will relabel the gauge potential as $A_{a}{}^{(\mu)}$
(without raising or lowering indices),
or simply as $A_{a}{}^{\mu}$ without the parenthesis.

Despite the fact that $A$ and $C$ transform in exactly the same way 
under general coordinate transformations,
it is not clear yet whether $A$ can be fully identified with $C$.
In particular, 
in pure GR,
not only the general coordinate transformation,
but also the rotations of local Lorentz frames are gauge symmetries.
We also need to check whether there are ghosts or tachyons
before we claim that 
the YM theory of the gauge symmetry of Sec.\ref{GaugeSymm}
can be interpreted as a gravity theory.
This will be the main issue to focus on below.

Nevertheless,
motivated by this potential identification,
we denote the covariant derivative as
\be
D_{a} = \hat{e}_{a}{}^{\mu}\del_{\mu},
\ee
where we used the notation
\be
\hat{e}_{a}{}^{\mu} \equiv 
\d_{a}^{\mu} + A_{a}{}^{\mu}.
\label{def-e-A}
\ee

The kinetic term of a scalar field is
\be
\eta^{ab}D_a \phi D_b \phi = \hat{g}^{\mu\nu} \del_{\mu} \phi \del_{\nu} \phi,
\ee
where the effective metric $\hat{g}_{\m\n}$ naturally arises.
It is defined by
\be
\hat{g}_{\mu\nu} = \hat{e}_{\m}{}^a \eta_{ab} \hat{e}_{\n}{}^b,
\label{metric}
\ee  
where $\hat{e}_{\m}{}^a$ is by definition the inverse of $\hat{e}_{a}{}^{\m}$.

\subsection{Field Strength vs Torsion}

The field strength of the non-Abelian gauge symmetry
constructed above is 
\be
F_{ab}(x) 
\equiv [D_a, D_b] 
= \del_{a} A_{b}(x) - \del_{b} A_{a}(x) + [A_{a}(x), A_{b}(x)].
\ee
In the representation (\ref{Teps}),
it is
\be
F_{ab}(x) = F_{ab}{}^{(\lam)}(x) \del_{\lam},
\ee
where
\be
F_{ab}{}^{(\lam)}(x) =
\hat{e}_a{}^{\mu}\del_{\mu}\hat{e}_b{}^{\lam} 
- \hat{e}_b{}^{\mu}\del_{\mu}\hat{e}_a{}^{\lam}.
\label{field-strength}
\ee

With the analogy between $\hat{e}_{\mu}{}^a$ and the vielbein $e_{\mu}{}^a$,
we define
\be
\hat{T}^{\lam}{}_{\mu\nu} \equiv
\hat{\Gamma}_{\mu\nu}^{\lam} - \hat{\Gamma}_{\nu\mu}^{\lam}.
\label{torsion}
\ee
It is the torsion for 
the Weitzenb\"{o}ck connection
\be
\hat{\Gamma}^{\lam}_{\mu\nu} \equiv
\hat{e}_a{}^{\lam}\del_{\mu}\hat{e}_{\nu}{}^a
\label{connection}
\ee
used in teleparallel gravity
when $\hat{e}_{\mu}{}^a$ is identified with the vielbein $e_{\mu}{}^a$.
The field strength and the ``torsion'' are essentially the same quantity:
\be
F_{ab}{}^{(\lam)}(x)
= - \hat{e}_a{}^{\mu} \hat{e}_b{}^{\nu} \hat{T}^{\lam}{}_{\mu\nu},
\ee
if we think of $\hat{e}_{\mu}{}^a$ and $\hat{e}_a{}^{\m}$
as the quantities used to switch between the two bases $\del_{\m}$ and $D_a$.

The ``connection'' (\ref{connection}) satisfies the relation
\be
D_{\mu} \hat{e}_{\nu}^a \equiv
\del_{\mu} \hat{e}_{\nu}{}^a - \hat{\G}_{\mu\nu}^{\lam} \hat{e}_{\lam}{}^a = 0,
\ee
and has zero ``curvature'':
\be
\hat{R} \equiv
d\hat{\Gamma} - \Gamma\wedge\Gamma = 0.
\ee

\subsection{Field-Dependent Killing Form}

An interesting feature of the algebra (\ref{commutator})
for space-time diffeomorphism is that 
the Killing form (invariant inner product)
has to be field-dependent.

For two elements of the Lie algebra
$T_f \equiv f^{\m}(x) \del_{\m}$ and 
$T_{f'} \equiv f'{}^{\n}(x) \del_{\n}$,
it is clear that the Killing form should be 
\be
\left< T_f | T_{f'} \right>
= \int d^D x \; \sqrt{\hat{g}} \; f^{\m}(x) \hat{g}_{\m\n}(x) f'{}^{\n}(x)
\label{Killing-1}
\ee
up to an overall normalization constant factor.
Here the measure $\sqrt{\hat{g}} = \det \hat{e}_{\mu}{}^a$
must be present to ensure that the integration is diffeomorphism-invariant. 
 
The Killing form can be slightly simplified by a change of basis.
Let us use the field-dependent basis $\{D_a\}$.
The Killing form for two generators
$T_f = f^a D_a$ and $T_{f'} = f'{}^b D_b$ is 
\be
\left< T_f | T_{f'} \right>
= \int d^D x \; \det \hat{e} \; f^{a}(x) \eta_{ab}(x) f'{}^{b}(x),
\label{Killing-2}
\ee
where $f^a \equiv f^{\mu} \hat{e}_{\mu}{}^a$
and similarly for $f'{}^b$.
(The factor of $\hat{g}_{\m\n}$ in (\ref{Killing-1}) is replaced by $\eta_{ab}$.)
In this basis,
the structure constants are field-dependent:
\be
[D_a, D_b] = F_{ab}{}^c D_c,
\label{DDD}
\ee
and the Jacobi identity (the consistency of the Lie algebra)
is equivalent to the Bianchi identity of the field strength.

\section{YM as Gravity}
\label{YM=GR}

\subsection{YM Action}

The YM action is given as the norm 
of the field strength:
\be
S_{YM} \equiv
\frac{1}{4\kappa^2} \left< F_{ab} | F^{ab} \right>
= \int d^D x \; \frac{\det \hat{e}}{4\kappa^2} F^{abc} F_{abc}.
\label{YM-action}
\ee
It is invariant under space-time diffeomorphism.
However,
it is not invariant under rotations of the local Lorentz frame:
\be
\hat{e}_{\m}{}^a(x) \rightarrow \hat{e}'_{\m}{}^a(x) = \om^a{}_b(x) \hat{e}_{\m}{}^b(x).
\label{rotation}
\ee

In the absence of the gauge symmetry of local Lorentz frame rotations,
the variable $\hat{e}_{\m}{}^a$ contains more degrees of freedom
than the genuine vielbein $e_{\m}{}^a$.
(This is why we have used a hat to distinguish it from the vielbein.)
The YM theory cannot be identified with pure GR.

To achieve a YM-like theory equivalent to Einstein's theory,
we should utilize the fact that 
the internal space index $a$ (for the basis $D_a$) can be contracted 
with the coordinate index $a$.
It allows us to introduce quadratic terms in addition to (\ref{YM-action}) in the action.
The most general quadratic action is
the superposition of three terms:
\be
S_{YM-like} = \int d^D x \; \frac{\det \hat{e}}{\kappa^2}
\left[ 
\frac{\lambda}{4} F^{abc} F_{abc}
+ \frac{\alpha}{4} F^{abc} F_{acb}
- \frac{\beta}{2} F^{ab}{}_{b} F_{ac}{}^{c}
\right].
\label{general-action-0}
\ee
The action remains the same if we simultaneously scale 
$\kappa^2, \lambda, \alpha, \beta$ by the same factor.
Up to this ambiguity, 
there is a unique choice of the parameters 
such that this action is invariant under local Lorentz rotations.

\subsection{Teleparallel Gravity}

The action (\ref{general-action-0}) is of a form resembling 
that of the teleparallel gravity,
which is equivalent to Einstein's theory \cite{review-tele}.

The teleparallel gravity has the interpretation as
the gauge theory of translational symmetry.
The gauge potential is essentially the vielbein:
\be
{\cal A}_{\m}{}^a \equiv e_{\m}{}^a - \d_{\m}{}^a,
\label{A=e-I}
\ee
where $\d_{\m}{}^a$ can be replaced by an arbitrary constant matrix.
The torsion $T^{\lam}{}_{\m\n}$ of 
the Weitzenb\"{o}ck connection
is essentially the Abelian field strength
\be
T^{a}{}_{\m\n} = \del_{\m} e_{\n}{}^a - \del_{\n} e_{\m}{}^a
= \del_{\m} {\cal A}_{\n}{}^a - \del_{\n} {\cal A}_{\m}{}^a.
\label{torsion-field-strength}
\ee

Despite the fact that the field strength (\ref{field-strength})
of the generalized gauge symmetry of diffeomorphism 
is related to this field strength (\ref{torsion-field-strength})
by a mere change of basis in the tangent space,
the gauge symmetries are totally different.
The gauge symmetry is Abelian in the traditional interpretation 
of teleparallel gravity,
while the diffeomorphism is of course non-Abelian.

The action of teleparallel gravity is
\bea
S_{TP} &=&
\int d^D x \; \frac{e}{2\kappa^2}
\left[ 
\frac{1}{4} T^{\lam\m\n} T_{\lam\m\n}
+ \frac{1}{2} T^{\m\n\lam} T_{\lam\n\m}
- T^{\lam\m}{}_{\lam} T^{\n}{}_{\m\n}
\right],
\label{action-TP}
\eea
where indices are raised or lowered 
using the metric $g_{\mu\nu} = e_{\m}{}^a e_{\n}{}^b \eta_{ab}$ 
and $e = \sqrt{g}$ stands for the determinant of $e_{\mu}^a$.
The Lagrangian of this action equals the Hilbert-Einstein Lagrangian 
up to a total derivative.
That is,
\be
{\cal L}_{TP} = \frac{e}{2\kappa^2} R^{(LC)} + \mbox{total derivatives},
\ee
where $R^{(LC)}$ is the scalar curvature for
the (torsion-free) Levi-Civita connection.
Even though the choices of connections are different,
the teleparallel gravity action and the Hilbert-Einstein action 
give exactly the same field equation for the metric
and so they are physically equivalent.

It is interesting that the inverse of $(I + {\cal A})$ (see (\ref{A=e-I}))
for the potential ${\cal A}$ of the Abelian group of translations
can be identified with $(I + A)$ for the gauge potential $A_{a}{}^{\m}$ of 
the non-Abelian gauge symmetry of general coordinate transformations.

The teleparallel gravity action (\ref{action-TP}) is equivalent to
the action (\ref{general-action-0}) for 
the choice of parameters $\lam = 1/2, \alpha = 1, \beta = 1$.
It is
\be
S_{TP} = \int d^D x \; \frac{\det \hat{e}}{2\kappa^2}
\left[ 
\frac{1}{4} F^{abc} F_{abc}
+ \frac{1}{2} F^{abc} F_{cba}
- F^{ab}{}_{b} F_{ac}{}^{c}
\right].
\label{action-TP-2}
\ee
The first term is the YM action (\ref{YM-action}).
The rest of the terms provide the unique combination so that 
the action is invariant under rotations of local Lorentz frames (\ref{rotation}).
The field $\hat{e}_{\m}{}^{a}$ can now be identified with the vielbein $e_{\m}{}^a$ in gravity,
and the modified YM theory is equivalent to GR.


\subsection{Metric, $B$-field and Dilaton}

While the action (\ref{action-TP-2}) is equivalent to pure GR,
we investigate the most general quadratic action (\ref{general-action-0}),
which can be equivalently put in the form
\be
S = \int d^D x \; \frac{\det\hat{e}}{2\kappa^2}
\left[ 
\frac{1}{2} F^{abc} F_{abc}
- \frac{\a}{12} (F^{abc}+F^{bca}+F^{cab})(F_{abc}+F_{bca}+F_{cab})
- \b F^{ab}{}_{b} F_{ac}{}^{c}
\right],
\label{general-action}
\ee
assuming that the coefficient of the YM term is non-zero.
It is invariant under general coordinate transformations
for arbitrary constants $\a, \b$.
(Compared with (\ref{general-action-0}),
$\lam = 1 - \a/2$.)
The case of teleparallel gravity (\ref{action-TP}) corresponds to the choice $\a = \b = 1$.

For generic values of $\a, \b$,
local rotations of Lorentz frames are no longer gauge symmetries.
With fewer gauge symmetries,
there are more physical degrees of freedom in the theory.
In $D$-dimensional space-time, 
the fundamental field $\hat{e}_a{}^{\m}$ has $D^2$ components.
When the rotation of local Lorentz frames is a gauge symmetry,
local Lorentz transformations identify $D(D-1)/2$ components 
of $\hat{e}_a{}^{\m}$ as gauge artifacts,
with the remaining $D(D+1)/2$ components of $\hat{e}_a{}^{\m}$ to be matched 
with the $D(D+1)/2$ independent components of the metric.

Tuning the values of $\a, \b$ slightly away from $1$,
we have $D(D-1)/2$ of the components that can no longer be gauged away.
The theory with generic values of $\a, \b$ 
is expected to contain more physical fields in addition to the metric.
For coefficients $\a, \b$ with values not too different from $1$,
the theory is expected to be a gravity theory including matter fields.
After all, 
the gauge symmetry of general coordinate transformation is always present.

In Einstein's theory of gravity,
for a fluctuation of the metric
\be
g_{\m\n} = \eta_{\m\n} + h_{\m\n} + \cdots,
\ee
one can choose the vielbein to be symmetric
\be
e_{\m a} = \eta_{\m a} + h_{\m a}/2 + \cdots
\ee
as a condition for the local Lorentz frame.
Note that in the perturbation theory
we are forced to mix the Latin and Greek indices
as the space-time is Minkowskian at the lowest order.
We will no longer distinguish the indices in the perturbative theory,
and use both sets of labells $a, b, c, \cdots$ and $\m, \n, \cdots$ at will.

We decompose the field $A_{\m a}$
into the symmetric part and the anti-symmetric part
\be
A_{ab} = (h_{ab} + B_{ab})/2,
\label{A=h+B}
\ee
where $h_{ab}$ is symmetric and $B_{ab}$ is anti-symmetric.
We identify $h_{ab}$ as the fluctuation of the metric,
and only the traceless part of $h_{ab}$ propagates in Einstein's theory.
When the rotation of the local Lorentz frame is not a gauge symmetry,
the trace part of $h_{ab}$ and the tensor $B_{ab}$ cannot be gauged away.

For the theory to be physically sensible,
one has to check that there are no ghosts or tachyons.
A necessary condition for linearized field equations of a rank-2 tensor field
to be free of ghosts and tachyonic modes is that 
the anti-symmetric part of the tensor field is decoupled from 
the symmetric part \cite{VanNieuwenhuizen:1973fi}.
The implication of this criterion to the general quadratic action (\ref{general-action})
can be easily derived as follows.
First, 
the cyclic combination 
\be
F_{abc} + F_{bca} + F_{cab} = 
(\del_a B_{bc} + \del_b B_{ca} + \del_c B_{ab}) + {\cal O}(A^2)
\label{FFF}
\ee
in the second term of the action (\ref{general-action})
involves only the anti-symmetric tensor field $B_{ab}$ at the linearized level.
(This was why we chose the peculiar form of the second term in the action
(\ref{general-action}).)
Hence its coefficient $\a$ has no effect on the coupling between $h_{ab}$ and $B_{ab}$.

The first and third terms in the action (\ref{general-action})
involve both $h_{ab}$ and $B_{ab}$,
and the relative magnitude of their coefficients should be fixed
to decouple $h_{ab}$ from $B_{ab}$.
Since we know that pure GR is free of ghosts and tachyons,
the ratio of these coefficients should be
identical to that of the teleparallel gravity action (\ref{action-TP}).
Consequently,
the parameter $\b$ should be fixed as
\cite{MuellerHoissen:1983vc}
\be
\b = 1.
\label{b=1}
\ee

It is still necessary to check that the kinetic terms are positive-definite.
The parameter $\a$ is constrained by \cite{MuellerHoissen:1983vc}
\be
\a < 1
\label{a<1}
\ee
for the kinetic term of the anti-symmetric field $B_{ab}$
to be positive-definite.
These two conditions (\ref{b=1}) and (\ref{a<1}) ensures that
the theory is ghost-free and tachyon-free at the free field level.

For this class of theories,
the propagating modes of the traceless part of $h_{ab}$ should be interpreted as the graviton,
and the trace part $h^a{}_a$ as the dilation.
There is also a rank-2 anti-symmetric field $B_{ab}$
whose gauge transformation at the lowest order is
\be
\d B_{ab} = \del_a \Lam_b - \del_b \Lam_a + {\cal O}(A).
\ee
One can define the covariant field strength of $B_{ab}$ as
$(F_{abc} + F_{bca} + F_{cab})$ (\ref{FFF}).

The dilaton can be viewed as a 0-form gauge potential.
In the Lorentz gauge
\be
\del_{\m} \hat{e}_a{}^{\m} = 0,
\label{Lorentz-gauge}
\ee
we have
\bea
F^{ab}{}_b &=&
\hat{e}^{ac}\del_c \hat{e}^b{}_b - \hat{e}^{bc}\del_c \hat{e}^{a}{}_b
\nn \\
&=&
\frac{D}{2} \del^a \hat\phi + {\cal O}(A^2),
\label{F-phi}
\eea
where 
\be
\hat\phi \equiv h^b{}_b/D,
\ee
so that $F_{ab}{}^b$ can be interpreted as the field strength of the dilaton.

\subsection{Comparison with Other Formulations of Gravity}

At first sight, 
the formulation of gravity outlined above may appear 
reminiscent to other known formulations of gravity as a gauge theory.
We have already commented above 
how the traditional interpretation of teleparallel gravity 
(as an Abelian gauge theory)
is different from our formulation.
There are also non-Abelian gauge theory formulations of gravity,
e.g. Chern-Simons gravity in 3D \cite{Witten:1988hc},
MacDowell-Mansouri gravity in 4D \cite{MacDowell:1977jt}
and higher dimensional generalizations \cite{Vasiliev:2001wa}.
In these theories,
the gauge potential is of the form
\be
A_{\m} = e_{\m}{}^a P_a + \omega_{\m}{}^{ab} J_{ab},
\label{usual}
\ee
where $P_a$ and $J_{ab}$ are generators of 
the local Poincare algebra $iso(D-1, 1)$ or the conformal algebra $so(D-1, 2)$.
The components of the gauge potential are the vielbein $e_{\m}{}^a$ 
and the spin connection $\om_{\m}{}^{ab}$.
In contract, 
the gauge potential in our theory is 
(in a certain representation)
\be
A_{a} = (\hat{e}_{a}{}^{\m}-\d^{\m}_a) \del_{\m},
\label{ours}
\ee
where $\hat{e}_{a}{}^{\m}$ might be identified 
with the {\em inverse} of the vielbein
if the action respects the local Lorentz symmetry.

Formally, 
if we identify the translation generator $P_a$ in (\ref{usual})
with the derivative $\del_{\m}$ in (\ref{ours}),
the gauge potential (\ref{usual}) resembles (\ref{ours}).
However, 
more precisely,
there are many important differences.
\begin{enumerate}
\item
The generators $P_a$ (and $J_{ab}$) in (\ref{usual}) 
commute with functions on the base space, 
while the derivative in (\ref{ours}) does not.
\item
The coefficient of $P_a$ is the vielbein,
and that of $\del_{\m}$ is the inverse vielbein 
in the special case of teleparallel gravity.
In general,
$\hat{e}_{\m}{}^a$ includes more 
degrees of freedom than the inverse vielbein.
\item
Eq. (\ref{usual}) is the usual potential 
associated with the ``gauging'' of a finite dimensional Lie group,
while (\ref{ours}) is not the potential for 
gauging any global symmetry.
\item
The field strength for (\ref{usual}) is given by the Riemann tensor, 
and that for (\ref{ours}) by the torsion of the Weitzenb\"{o}ck connection.
A priori they are not related in any simple way
as the Weitzenb\"{o}ck connection is not invariant under 
local Lorentz transformations while the Riemann tensor is.
\item
Despite the fact that the Lagrangians for gravity 
are quadratic in the field strengths for both potentials 
(\ref{usual}) and (\ref{ours}).
The former gives the Hilbert-Einstein action.
Even in the special case of telelparallel gravity,
the latter differs by a total derivative.
\end{enumerate}

Supergravity theories are constructed \cite{SUGRA}
based on the YM-like theory for the gauge potential (\ref{usual}).
It will be interesting to consider the supersymmetrization 
of the gauge symmetry of diffeomorphism
and to derive the supergravity theory 
as an alternative formulation of supergravity.
We leave this project for future publications.

\section{Scattering Amplitudes}
\label{ScattAmp}

In recent years, 
there has been amazing progress in 
the techniques of calculating scattering amplitudes,
as well the understanding of their structures.
Among them, 
a very interesting and mysterious structure
is the connection between YM theory and gravity
through the so-called double-copy procedure,
which utilizes the color-kinematics duality 
(also known as the BCJ duality) \cite{BCJ}.
According to Ref.\cite{BCJ},
certain gravity theories are double copies of YM theories:
the scattering amplitudes of gravity theories 
can be obtained from those of YM theories 
with color factors replaced by kinematic factors.
In many cases this connection can find its origin in 
the open-closed string duality (the KLT duality \cite{KLT}), 
although there are also other cases in which 
the string-theory origin is absent at this moment.

A complete off-shell field-theoretic explanation 
of this connection between gravity and YM theories,
which applies only to on-shell amplitudes,
may not be possible.
But it is desirable to understand how much of 
the on-shell miracle can be understood in an off-shell theory.
Earlier efforts in this direction include 
Refs. \cite{BjerrumBohr:2012mg,Monteiro:2014cda}.
We propose that 
the formulation of gravity as a YM theory we constructed above
may shed some new light on this problem.

\subsection{Heuristic Explanation}
\label{heuristic}

Let us first re-examine the double-copy procedure
to illustrate our idea. 
As the simplest example,
for the pure YM theory,
the color-ordered 3-point amplitude at tree-level is
\be
f_{abc} \, n^{(3)}_{\lam\m\n}(p, q, -(p+q))
\label{fn}
\ee
where $f_{abc}$ is the structure constant 
and $n^{(3)}_{\lam\m\n}(p, q, -(p+q))$ the kinematic factor
\be
n^{(3)}_{\lam\m\n}(p, q, r) \equiv
(p-q)_{\n}\eta_{\lam\m} + (q-r)_{\lam}\eta_{\m\n} + (r-p)_{\m}\eta_{\n\lam}.
\label{n}
\ee
Here $(p, q, r)$ are the momenta of the 3 external legs, 
and $\lam, \m, \n$ are Lorentz indices labelling 
the polarizations of the vector fields.
The origin of this factor (\ref{n}) is the 3-point vertex 
\be
[A_{\m}, A_{\n}]_c \, F_{(0)}^{\m\n c} =
f_{abc} \, A_{\m}^a A_{\n}^b \, (\del^{\m} A^{\n c} - \del^{\n} A^{\m c})
\label{fAAF}
\ee
in the YM Lagrangian,
with cyclic permutations of the three factors of $A$ 
contracted with three external legs, 
assuming that the basis of the Lie algebra is chosen such that 
$f_{abc} = f_{bca} = f_{cab}$.
Here $F_{(0)}^{\m\n c} \equiv \del^{\m} A^{\n c} - \del^{\n} A^{\m c}$
is the field strength at the lowest order.

The double-copy procedure states that 
the replacement of $f_{abc}$ by $n^{(3)}_{\lam\m\n}(p, q, -(p+q))$
gives the 3-point amplitude of the corresponding gravity theory.
In other words,
if there is a Lie algebra with indices $a = (\lam, p)$
and structure constants $f_{abc}$ given by $n^{(3)}_{\lam\m\n}(p, q, -(p+q))$,
the YM theory would agree with GR at least for 3-point amplitudes.
Yet one can check that
this choice of structure constants does not satisfy the Jacobi identity.

The color-kinematics duality and the double-copy procedure 
applies to all higher-point amplitudes.
For 4-point amplitudes at the tree level,
the kinematic factor of color-ordered amplitudes is
\footnote{
More explicitly \cite{Zhu:1980sz},
\bea
n_s^{(4)} &=& \eps_1^{\lam}\eps_2^{\mu}\eps_3^{\nu}\eps_4^{\s}
n^{(3)}_{\s\n\r}(-p_4, -p_3, p_3+p_4)n^{(3)}_{\lam\m}{}^{\r}(p_1,p_2,-p_1-p_2)
+ m_s^{(4)},
\eea
where $p_i^{\m}, \eps_i^{\m}$ ($i = 1, 2, 3, 4$) 
are the momenta and polarization vectors of the external legs,
$s = (p_1+p_2)^2$ and
\be
m_s^{(4)} \equiv 
s[(\eps_1\cdot\eps_4)(\eps_2\cdot\eps_3)-(\eps_1\cdot\eps_3)(\eps_2\cdot\eps_4)].
\ee
(The choice of $m_s^{(4)}$ is not unique.)
The other two kinematic factors $n^{(4)}_t, n^{(4)}_u$ can be obtained 
by permutations of external legs.
},
skematically,
\be
n_i^{(4)} = n_i^{(3)}n_i^{(3)} + m_i^{(4)}
\qquad
(i = s, t, u),
\label{nn+m}
\ee
where the first term on the right hand side 
is the contribution from 3-point vertices, 
and the second term from the 4-point interaction in the YM theory.
(The index values $s, t, u$ are labels for the $s$, $t$ and $u$ channels
of the Feynman diagrams for tree-level 4-point scattering amplitudes.)
The kinematic factors $n_i^{(4)}$ satisfy a linear relation
\be
n_s^{(4)} + n_t^{(4)} + n_u^{(4)} = 0
\ee
analogous to the Jacobi identity for the color factor
(which is quadratic in the structure constants).
The relation above would be equivalent to the Jacobi identify 
for $n^{(3)}$ interpreted as structure constants
if the terms $m_i^{(4)}$ were absent.
The presence of $m_i^{(4)}$ is the evidence that 
$n^{(3)}$ cannot be used as structure constants.

Often the relation between YM and GR indicated by the double-copy procedure
is symbolically represented as (YM)${}^2 = $(GR).
This expression is actually misleading, 
because neither the color factors or the propagators are squared 
in the gravity theory.
Instead, 
the identification of color factors with kinematic factors
(e.g. $f_{abc}$ with $n_{\lam\m\n}(p, q, -(p+q))$ for tree-level 3-point amplitudes)
identifies YM directly with GR.
It is more appropriate to use (YM)${}^{\prime}$ = GR 
as the symbolic representation of this connection.
The prime on (YM) indicates the modification of YM theory 
by the replacement of color factors by kinematic factors.

Since the color factors are composed of structure constants of the gauge group,
we are naturally led to consider the possibility of gauge symmetries 
with structure constants involving kinematic factors.
This was precisely what we did in Sec.\ref{GaugeSymm},
which led to new formulations of gravity theories 
as generalized YM theories in Sec.\ref{YM}

Apparently, 
the hope for a direct matching between structure constants 
and the kinematic factors is too naive.
First, 
as the 3-point amplitude is defined on-shell,
the structure constant can be different from 
$n_{\lam\m\n}(p, q, -(p+q))$ when it is off-shell.
Secondly,
even if the 3-point amplitudes agree with structure constant, 
it is not clear if higher-point amplitudes will automatically agree 
with the corresponding color factors,
as there will be different on-shell conditions at work.
(In fact, eq.(\ref{nn+m}) says that 
the structure constants for 4-point amplitudes 
are not to be given by the structure constants for 3-point amplitudes
due to the extra term $m^{(4)}_i$.)
In general, 
as the BCJ duality only holds on-shell,
the correspondence between structure constants and kinematic factors 
in $n$-point amplitudes is different for different $n$,
and it is unclear if there exists an off-shell generalization.
It is also highly nontrivial how such on-shell correspondences 
can be implemented efficiently in a field-theoretic approach.
Hence we leave the search for a field-theoretic proof of
the validity of the double-copy procedure for future works.
Nevertheless,
the color-kinematics duality and double-copy procedure motivate us
to explore YM theories with Lie algebras involving kinematic factors,
and it does lead to a connection between 
YM theories and gravity theories 
at the level of Lagrangians as we have shown in Sec.\ref{YM}.

It will be interesting to see whether 
the calculation of scattering amplitudes is simplified
in this YM-like formulation of gravity,
compared with the calculation based on the Hilbert-Einstein action.
It will be even more interesting to see
if we are getting closer to the simplified on-shell results
obtained via the BCJ duality.

To gain some intuition about how far or close
our theory is to the concise results of the BCJ duality,
let us comment on the skematic structure of the 3-point scattering amplitude (\ref{fn}).
The kinematic factor $n$ comes from the factor $F_{(0)}$,
as the only object involving derivatives
in the cubic term $f AA F_{(0)}$ (\ref{fAAF}) of the YM action.
This suggests that,
to replace the color factor $f$ by $n$, 
we should have $f \sim F_{(0)}$ to the lowest order in $A$.
But this is precisely what we have:
the structure constant (\ref{DDD}) in the basis of $D_a$ is the field strength $F$!

In the following,
we will compute more carefully the 3-point vertices of the YM-like formulation of gravity,
and see that the calculation is much simpler 
than the calculation based on the Hilbert-Einstein action
(which involves around 100 terms).
We leave higher-point scattering amplitudes for the future.

Incidentally,
although the Chern-Simons theory in 3D \cite{Witten:1988hc} 
and the MacDowell-Mansouri theory in 4D \cite{MacDowell:1977jt} 
are also YM-type formulations of gravity,
they are first order formulations of GR.
One has to first solve the connection in terms of the vielbein 
before calculating any scattering amplitudes of gravitons.
The calculation in those theories is not simpler than
a direct computation from the Hilbert-Einstein action.

\subsection{Perturbative Expansion in $A$}

In this section, 
we focus on the 3-point vertices relevant for the 3-graviton scattering.
We consider 3-point vertices of the action (\ref{general-action}) for 
the traceless part of $h_{\m\n}$.

First, 
using (\ref{FFF}),
one can easily see that 
the second term in the action (\ref{general-action}) is
\bea
\frac{\a}{12} (F^{abc}+F^{bca}+F^{cab})(F_{abc}+F_{bca}+F_{cab})
&=&
\frac{\a}{3} H^{(0)abc} H^{(0)}_{abc} + {\cal O}(H^{(0)} A^2) + {\cal O}(A^4),
\eea
where $H^{(0)}_{abc}$ is the field strength of the anti-symmetric tensor field $B_{ab}$
defined at the lowest order:
\be
H^{(0)}_{abc} \equiv \del_a B_{bc} + \del_b B_{ca} + \del_c B_{ab}.
\ee
A vertex operator involving only external legs of $h$ appears 
at ${\cal O}(A^4)$ or higher.
The 3-point vertices of $h_{\m\n}$ is thus independent of the parameter $\a$.

Secondly,
the third term in (\ref{general-action}) is the square of
\bea
F_{ab}{}^b &=& 
\del_a A_b{}^b - \del_b A_a{}^b + {\cal O}(A^2),
\eea
where the first term involves the trace of $h_{\m\n}$,
and the second term vanishes if we impose the gauge-fixing condition
\be
\del^b h_{ab} = 0
\label{dh=0}
\ee
for the graviton field.
As a result, 
in this gauge (\ref{dh=0}),
the third term of the action (\ref{general-action})
is also irrelevant to the 3-point vertex for the traceless part of $h_{\m\n}$.

Furthermore,
the overall measure of integration is
\be
\det e = 1 + h_a{}^a + {\cal O}(A^2),
\ee
which is also irrelevant for our consideration.

The 3-point vertices for the graviton can thus only 
come from the YM Lagrangian,
and there are only two terms
\be
{\cal L}^{(3)} \equiv
F^{abc}_{(0)}([A_a, A_b]_c - F^{(0)}_{ab}{}^{d}A_{dc}),
\label{3-point-1}
\ee
where 
\be
[A_a, A_b]_c \equiv A_a{}^d\del_d A_{bc} - A_b{}^d\del_d A_{ac}
\ee
and $F^{(0)}_{abc}$ is the field strength at the zero-th order
\be
F^{(0)}_{abc} \equiv \del_a A_{bc} - \del_b A_{ac}.
\ee
The first term of (\ref{3-point-1}) comes from 
the Lie algebra structure of this generalized YM theory.
The second term arises due to the field-dependent inner-product of the Lie algebra.

Near the end of Sec.\ref{heuristic},
we discussed how the formulation of gravity
as a generalized YM theory can heuristically explain 
the double-copy procedure for 3-point amplitudes at tree level.
Eq.(\ref{3-point-1}) is the exact expression of 
the heuristic expression $f AAF_{(0)} \sim F_{(0)} A F_{(0)}$ there.
It may seem that there is a small discrepancy between
$F_{(0)}([A,A]+AF_{(0)})$ (\ref{3-point-1}) and $F_{(0)}AF_{(0)}$.
But recall that the structure constant $f_{abc}$ is assumed to 
be cyclic in the double-copy procedure
(and in our heuristic discussion in Sec.\ref{heuristic}),
while $F_{(0)}^{abc}$ is not.
The exact expression (\ref{3-point-1}) is in fact of the form 
of $F_{(0)}AF_{(0)}$ but additional terms that 
have some of the indices permuted.
In Sec.\ref{pert-hatA} below,
we will see a simpler and more direct match with
the discussions in Sec.\ref{heuristic}.

The 3-point vertices for the gravitons are therefore
\be
(\del^a h^{bc}-\del^b h^{ac})
\left[h_{ad}\del^d h_{bc}-h_{bd}\del^d h_{ac}
- (\del_a h_b{}^d - \del_b h_a{}^d)h_{dc}\right].
\ee
This is already a very simple expression,
especially if we compare it with the expression obtained 
from the Hilbert-Einstein action.
But the expression can be even further simplified 
as we will shown below.

\subsection{Perturbative Expansion in $\hat{A}$}
\label{pert-hatA}

It is more economic, 
at least at the lowest order,
to use the variable $\hat{A}_{\m a}$ defined by
\be
\hat{e}_{\m}{}^a = \d_{\m}^a + \hat{A}_{\m}{}^{a},
\ee
where $\hat{e}_{\m}{}^a$ is the inverse of $\hat{e}_a{}^{\m}$ (\ref{def-e-A}).
The new variable $\hat{A}_{\m}{}^a$ is merely a field redefinition of $A_a{}^{\m}$.
They are related via
\be
A_a{}^{\m} = \hat{e}_a{}^{\m} - \d_a^{\m}
= - \hat{A}_a{}^{\m} + \hat{A}_a{}^{\n}\hat{A}_{\n}{}^{\m} + {\cal O}(A^3).
\ee
Therefore
\be
h_{ab} \simeq - \hat{h}_{ab} + \cdots.
\ee
Up to sign, 
the physical (on-shell) amplitudes of $h_{ab}$ and $\hat{h}_{ab}$ should agree.
As it is suggested by the notation,
we have decomposed $\hat{A}$ as
\be
\hat{A}_{ab} = \hat{h}_{ab}/2 + \hat{B}_{ab},
\ee
where $\hat{h}_{ab}$ is symmetric
and $\hat{B}_{ab}$ is anti-symmetric.
The trace part of $\hat{h}_{ab}$ is denoted
\footnote{
As we have learned from pure GR, 
the trace part of the fluctuation of the metric 
is not a physical propagating mode.
Hence we should identify the trace part of $\hat{A}_{ab}$
(and $A_{ab}$)
as an independent scalar field.
}
\be
\hat{\phi} \equiv \frac{1}{D} \hat{h}_a{}^a.
\ee

The effective metric (\ref{metric}) is
\be
\hat{g}_{\m\n} = \hat{e}_{\m}{}^a\eta_{ab}\hat{e}_{\n}{}^b
\simeq \eta_{\m\n} + \hat{h}_{ab} + 2 \eta_{ab} \hat{\phi} + {\cal O}(A^2).
\ee
We also have
\be
\mbox{tr}\hat{e} \equiv
\hat{e}_{\m}{}^{a}\d_a^{\m} = D (1 + \hat{\phi} + {\cal O}(A^2)),
\ee
and thus
\be
\det \hat{e} = 1 + D\hat{\phi} + {\cal O}(A^2).
\ee
One can ignore the integration measure $\det\hat{e}$ 
when the 3-point vertex under consideration 
does not involve $\hat{\phi}$ as an external leg.

Expanding the field strength in powers of $\hat{A}$, 
we have
\bea
F_{abc} 
&=&
- \hat{e}_a{}^{\m}\hat{e}_b{}^{\n}(\del_{\m}\hat{A}_{\n c} - \del_{\n}\hat{A}_{\m c}),
\eea

Then,
\bea
\frac{1}{2}F_{abc}F^{abc} &=&
\frac{1}{2}
\hat{F}^{(0)}_{abc}\hat{F}^{(0)abc}
- 2 \hat{F}^{(0)abc}\hat{A}_a{}^{d}\hat{F}^{(0)}_{dbc} + {\cal O}(A^4),
\label{F2}
\eea
where
\be
\hat{F}^{(0)}_{\m\n a} \equiv
\del_{\m}\hat{A}_{\n a} - \del_{\n}\hat{A}_{\m a}.
\ee

Similar to the perturbative expansion in terms of $A$,
the second and third terms in the action (\ref{general-action}) 
do not contribute to 3-point interactions of
the traceless part of $h_{ab}$.

It is remarkable that for this action (\ref{general-action})
there is a single 3-point vertex for graviton interaction
(the second term in (\ref{F2}))
\be
\hat{\cal L}^{(3)} = - 2 \hat{F}^{(0)abc}\hat{A}_a{}^{d}\hat{F}^{(0)}_{dbc}.
\ee
This is precisely of the form $F_{(0)}AF_{(0)}$ needed to explain 
the double-copy procedure at tree level 
as we discussed in Sec.\ref{heuristic}.
This is of course also a significant simplification compared with
the usual expression of GR.

\section{Higher-Form Gauge Symmetries}
\label{HigherGauge}

In the above we have focused on the gauge symmetries 
with 1-form potentials and 0-form gauge parameters.
We can also apply the same notion of generalization to 
gauge symmetries with higher-form gauge potentials.
The basic ideas of our generalization are the following:
\begin{enumerate}
\item
\label{g1}
The symmetry generators do not have to be factorizable in the form
\be
T(\{f, a\}) = \sum_n f_n(x) T_{a_n},
\label{factorizable}
\ee
where $f_n(x)$'s are functions on the base space 
and $T_a$'s are elements of a finite-dimensional Lie algebra.
\item
\label{g2}
Even if the symmetry generators are formally factorable in the form (\ref{factorizable}),
the objects $T_a$'s do not have to commute with base-space functions.
\end{enumerate} 
If we take a given principal bundle as a classical manifold
and deform its algebra of functions so that it becomes noncommutative,
in general, 
this noncommutative space is not
the tensor product of a group and a noncommutative base space.
This is a way to construct examples of the comment \ref{g1} above.

As an example of the comment \ref{g2},
even though
the gauge symmetry constructed in Sec.\ref{GaugeSymm}
has generators of the form (\ref{factorizable}) 
with $T_a = \del_a$,
these $T_a$'s should not be interpreted as generators of 
a finite-dimensional Lie algebra
(otherwise the Lie algebra is Abelian)
and they do not commute with space-time functions.

For higher-form gauge symmetries,
only the Abelian case is well understood.
There is no consensus on the definition of 
non-Abelian higher-form gauge symmetries,
\footnote{
The mathematical structure for the symmetry of a 2-form gauge potential
is called a non-Abelian gerbe.
But there are different versions of its definition.}
and concrete examples are scarce. 
Due to this reason,
our discussion below cannot be very precise.
It is commonly speculated that 
there is something analogous to the Lie algebra 
whose elements replace $T_a$ in the factorized formula (\ref{factorizable})
of a gauge transformation generator.
The analogue of our generalization of gauge symmetry
for higher-form gauge symmetries
is then referring to a violation of that factorization.

The Nambu-Poisson gauge theory \cite{M5-C,NP} is one of the few examples
of non-Abelian higher-form gauge theories.
It was used to describe an M5-brane in a large $C$-field background.
It can be viewed as the covariant lift of 
the Poisson limit of the noncommutative gauge symmetry 
for a D4-brane in large $B$-field background to a higher dimension.
The Nambu-Poisson gauge symmetry 
has a 2-form potential with 1-form transformation parameters.
Its gauge group is the non-Abelian group of volume-preserving-diffeomorphisms.
It can be viewed as an example of generalized gauge symmetry 
for higher form potentials.

Incidentally,
the fact that the gauge algebra involves kinematic factors 
is also the reason why it is possible for higher-form gauge symmetries to be non-Abelian.
Higher-form global symmetries are always Abelian \cite{Gaiotto:2014kfa}.
Hence an ordinary procedure of ``gauging'' the global symmetry 
by introducing space-time dependence to the generators 
in a way analogous to eq.(\ref{factorizable})
can never result in a non-Abelian gauge symmetry
(unless the space-time coordinates are noncommutative).
Conversely, 
for a non-Abelian higher-form gauge symmetry,
when the transformation parameters are restricted to be constant,
all kinematic factors become trivial, 
and the symmetry algebra becomes Abelian.
The Nambu-Poisson gauge symmetry is clearly an example of this fact.
The noncommutative $U(1)$ gauge symmetry is the lower-form analogue.

Another example of non-Abelian gauge symmetry with a 2-form gauge potential
is the low energy effective theory for 
multiple M5-branes proposed in Ref.\cite{MM5,Ho:2012nt}.
The M5-branes are compactified on a circle,
and the gauge transformation laws distinguish zero-modes from KK modes.
The distinct treatment on zero-modes and KK modes 
can be viewed as a dependence on the kinematic factor
(whether the momentum is zero or not),
and so it is also an example of the generalized gauge theory for higher forms.

There are other examples of non-Abelian gauge symmetry
with higher-form gauge potentials \cite{2form},
in addition those mentioned above.
It will be interesting to explore further how the idea 
promoted above on generalized gauge symmetry 
will help the construction of a mathematical framework
for non-Abelian higher-form gauge theories.

\section*{Acknowledgement}

The author would like to thank 
Chong-Sun Chu, Kazuo Hosomichi, Yu-Tin Huang, Takeo Inami
for their interest and discussions.
This work is supported in part by
the National Science Council, Taiwan, R.O.C.
and by the National Taiwan University.


\end{CJK} 
\end{document}